\documentstyle[12pt]{article}
\def\beq{\begin{equation}}
\def\eeq{\end{equation}}
\def\bea{\begin{eqnarray}}
\def\eea{\end{eqnarray}}
\def\bem{\begin{math}}
\def\eem{\end{math}}
\def\bit{\begin{itemize}}
\def\eit{\end{itemize}}
\def\bla{\begin{flushright}}
\def\ela{\end{flushright}}
\def\qq2{$Q^2$}               
\def\aa1{$A_1(x,Q^2)$}        
\def\ff1{$F_1(x,Q^2)$}        
\def\gg1{$g_1(x,Q^2)$}        
\begin{document}
\vskip 3cm
\begin{center}
{\Large{\bf The $Q^2$ dependence of the measured asymmetry $A_1$:
the test of the Bjorken sum rule}}
\end{center}
\vskip 1.7cm
\begin{center}
{\bf A.V.Kotikov}
\footnote{
~E-mail: kotikov@sunse.jinr.dubna.su}
and  {\bf D.V.Peshekhonov}
\footnote{~E-mail:
peshehon@sunse.jinr.dubna.su}
\\
{\it Particle Physics Laboratory\\
 Joint Institute for Nuclear Research\\
141980 Dubna, Russia.}
\end{center}
\vskip 3cm
{\large{\bf Abstract}}\\
\vskip .5cm \hskip -.56cm
We analyse the proton and deutron data on spin dependent asymmetry
~\aa1 supposing the DIS structure functions $g_1(x,Q^2)$
and $F_3(x,Q^2)$ have the similar $Q^2$-dependence.
As a result, we have obtained that
$\Gamma_1^p - \Gamma_1^n = 0.190 \pm 0.038$ at
$Q^2= 10~{\rm GeV}^2$ and $\Gamma_1^p - \Gamma_1^n = 0.165 \pm 0.026$ at
$Q^2= 3~{\rm GeV}^2$, what is in the best agreement with the Bjorken
sum rule predictions.\\
PACS number(s): 13.60.Hb, 11.55.Hx, 13.88.+e \\
\newpage
\hskip -.56cm
An experimental study of the nucleon spin structure is realized by the
measuring of asymmetry $A_1(x,Q^2) = g_1(x,Q^2) / F_1(x,Q^2)$.
The best known theoretical predictions on spin dependent structure
function $g_1(x,Q^2)$ of the nucleon were made by Bjorken \cite{Bj} and
Ellis and Jaffe \cite{EJ} for the so called {\it first moment value}
$\Gamma_1 = \int_0^1 g_1(x) dx$.\\
The calculation of the $\Gamma_1$ value requires the knowledge of
structure function $g_1$ at the same $Q^2$ in the whole $x$ range.
Experimentally the asymmetry $A_1$ is measuring at different values of $Q^2$
for different $x$ bins.
An accuracy of the past and modern experiments \cite{SMCp,E143p}
allows to analyze data in the assumption \cite{EK93}
that asymmetry \aa1 is \qq2
independent (i.e. the structure functions $g_1$ and $F_1$ have the same $Q^2$
dependence).
However, this assumption is not theoretically warranted (see
discussions in \cite{AEL,GS95,GRSV}); the different $Q^2$ dependence
of the structure functions $g_1(x,Q^2)$ and $F_1(x,Q^2)$ is expected
due to the difference in polarized and unpolarized spliting functions
(except for the leading order quark-quark one). Thus, in view of
forthcoming more precise data it is important to add the $Q^2$
dependence of the asymmetry.
\vskip .4cm \hskip -.56 cm
This article is based on our observation that the $Q^2$ dependence of
spin dependent and spin average structure functions $g_1$ and $F_3$ is
very similar in a wide $x$ range:
$10^{-2} < x < 1$. At the small $x$ region ($x<10^{-2}$ it could be not
true (see \cite{AEL,smallx}), but most of the existed data were measured out
off that range.\\

\hskip -.56cm
Lets consider the nonsinglet (NS) $Q^2$ evolution of structure functions
$F_1,~g_1$ and $ F_3$. The DGLAP equation for the NS part of these functions can be presented as\footnote{We use $\alpha(Q^2)= \alpha_s(Q^2)/{4 \pi}$ .} :\\
\bea
{dg_1^{NS}(x,Q^2) \over dlnQ^2} &=& -{1 \over 2} \gamma_{NS}^-(x, \alpha)
\times g_1^{NS}(x,Q^2),
\nonumber \\
{dF_1^{NS}(x,Q^2) \over dlnQ^2} &=& -{1 \over 2} \gamma_{NS}^+(x, \alpha)
\times F_1^{NS}(x,Q^2),
\label{1} \\
{dF_3(x,Q^2) \over dlnQ^2} &=& -{1 \over 2} \gamma_{NS}^-(x, \alpha)
\times F_3(x,Q^2),
\nonumber
\eea
where symbol $\times$ means the Mellin convolution. The spliting
functions $\gamma^{\pm}_{NS}$ are the reverse Mellin transforms of the
anomalous dimensions $\gamma^{\pm}_{NS}(n, \alpha)= \alpha
\gamma^{(0)}(n)_{NS}  +
\alpha^2 \gamma^{\pm (1)}_{NS}(n) + O(\alpha^3)$ and the Wilson
coefficients\footnote{Because we consider here the
structure functions themselves
but not the parton distributions. Note that $b^{+}_{NS}(n)$ and
$b^{-}_{NS}(n)$ have
more standard definition as $b_{1,NS}(n) = b_{2,NS}(n) - b_{L,NS}(n)$ and
$b_{3,NS}(n)$.}
$\alpha b^{\pm}(n) + O(\alpha^2)$ :
 \begin{eqnarray}
\gamma^{ \pm}_{NS}(x,\alpha) ~=~ \alpha
  \gamma^{(0)}_{NS}(x) + \alpha^2 \biggl(
  \gamma^{\pm (1)}_{NS}(x) +
  2\beta_0 b^{\pm}(x)  \biggr) + O(\alpha ^3),
\label{2}  \end{eqnarray}
where $\beta (\alpha)= - \alpha^2 \beta_0 -
 \alpha^3 \beta_1 + O(\alpha^4)$ is QCD $\beta$-function.\\
The above mentioned Mellin transforms mean that
 \begin{eqnarray}
f(n,Q^2) ~=~ \int^1_0 dx x^{n-1} f(x,Q^2),
\label{3}  \end{eqnarray}
where $ f= \{\gamma^{(0)}_{NS}, \gamma^{\pm (1)}_{NS}, b^{\pm}_{NS},
\gamma^{(k)}_{ij}, \gamma^{* (k)}_{ij}, b_{i} \mbox{ and } b^{*}_{i} \} $
with $k = 1,2$ and $\{i,j \} = \{S,G \}$. \\
Eqs. (1) show the $Q^2$ dependence of NS parts of $g_1$ and $F_3$ is the same
(at least in first two orders of the perturbative QCD \cite{Kodaira})
and differs from $F_1$ already in the first subleading order
$\biggl(\gamma^{+(1)}_{NS} \neq \gamma^{-(1)}_{NS}$ \cite{RoSa} and
$b^+_{NS} - b^-_{NS} = (8/3)x(1-x)\biggr)$.\\
For the singlet parts of $g_1$ and $F_1$ evolution equations are :\\
\bea
{dg_1^S(x,Q^2) \over dlnQ^2} &=& -{1 \over 2} \biggl[
\gamma_{SS}^{*}(x, \alpha) \times g_1^S(x,Q^2) +
\gamma_{SG}^{*}(x, \alpha) \times \Delta G(x,Q^2) \biggr],
\nonumber
\\
{dF_1^S(x,Q^2) \over dlnQ^2} &=& -{1 \over 2} \biggl[
\gamma_{SS}(x, \alpha) \times F_1^S(x,Q^2) +
\gamma_{SG}(x, \alpha) \times G(x,Q^2) \biggr],
\label{4}
\eea
where
 \bea
\gamma_{SS}(x,\alpha) &=& \alpha
  \gamma^{(0)}_{SS}(x) + \alpha^2 \biggl(
  \gamma^{(1)}_{SS}(x) +  b_G(x) \times \gamma^{(0)}_{GS}(x) +
  2\beta_0 b_S(x)  \biggr)
+ O(\alpha ^3),
\nonumber  \\
\gamma_{SG}(x,\alpha) &=& \frac{e}{f} \biggl[ \alpha
  \gamma^{(0)}_{SG}(x) + \alpha^2 \biggl(
  \gamma^{(1)}_{SG}(x) + b_G(x) \times \bigl(
  \gamma^{(0)}_{GG}(x) - \gamma^{(0)}_{SS}(x) \bigr)
+ 2\beta_0 b_G(x)
\nonumber \\
&+&  b_S(x) \times \gamma^{(0)}_{SG}(x)
\biggr)
  \biggl] + O(\alpha ^3) \nonumber  \eea
where $e = \sum_i^f e^2_i$ is
the sum of charge squares of $f$ active quarks.
The equations for polarized anomalous dimensions
$\gamma_{SS}^{*}(x, \alpha)$ and $\gamma_{SG}^{*}(x, \alpha)$
are similar. They can be obtained by replacing
$\gamma^{(0)}_{SG}(x) \to \gamma^{*(0)}_{SG}(x)$,
$\gamma^{(1)}_{Si}(x) \to \gamma^{*(1)}_{Si}(x)$ and $b_i(x) \to
b^*_i(x)$ ($i=\{S,G \}$).\\

Note here the gluon term is not negligible for $F_1$ at $x < 0.3$ but for
$g_1$ we can neglect them for $x>0.01$
\cite{GS95,GRSV}.
The value $b^*_s(x)$ ($b_s(x)$) coincides with $b^-(x)$ ($b^+(x)$).
The difference between
$\gamma_{NS}^{-(1)}$ and $\gamma_{SS}^{* (1)}+  b^*_G(x) \times
\gamma^{(0)}_{GS}(x)$  is negligible because it does not contain
a power singularity at $x \to 0$
(i.e. a singularity
at $n \to 1$ in momentum space).
Moreover, it decreases as $O(1-x)$ at $x \to 1$
\cite{MeNe}.
Contrary to this, the difference between
$\gamma_{SS}^{(1)}+  b_G(x) \times \gamma^{(0)}_{GS}(x)$ and
$\gamma_{SS}^{* (1)}+  b^*_G(x) \times \gamma^{(0)}_{GS}(x)$
contains the power singularity at $x \to 0$ (see for example
\cite{Kodaira}).\\

The analysis discussed above allows us to conclude the function $A_1^*$ :
\bea
A_1^*(x) = {g_1(x,Q^2) \over F_3(x,Q^2)}
\label{4.1}
\eea
should be practically $Q^2$ independent at $x>0.01$.

The r.h.s. of Eqs.(\ref{1}) and (\ref{4}) contain integrals of
structure functions and, hence, the approximate validity of (\ref{4.1})
should be observed only for the similar $x$-dependence of $g_1(x,Q^2)$
and $F_3(x,Q^2)$ at fixed $Q^2$. But it is the case
 (see \cite{BoSo} at $Q^2 = 3 GeV^2$, for example). \\
The asymmetry $A_1$ at $Q^2=<Q^2>$ can be defined than as :
\bea
A_1(x_i,<Q^2>) =  {F_3(x_i,<Q^2>) \over F_3(x_i,Q^2_i)} \cdot
{F_1(x_i,Q^2_i) \over F_1(x_i,<Q^2>)} \cdot A_1(x_i,Q^2_i),
\label{5}
\eea
where $x_i$ ($Q^2_i$) means an experimentally measured value of $x$ ($Q^2$).\\
\vskip .4cm \hskip -.56cm
We use SMC and E143 proton and deuteron data on asymmetry $A_1(x,Q^2)$
\cite{SMCp, E143p}. To get $F_1(x,Q^2)$ we take NMC parametrization
of $F_2(x,Q^2)$ \cite{NMC}
and SLAC parametrization of $R(x,Q^2)$ \cite{SLAC} ($F_1 \equiv
F_2/2x[1+R]$).
To get the values of $F_3(x,Q^2)$ we parametrize the CCFR data \cite{CCFR} as
a function of $x$ and $Q^2$ (see the parametrization in Appendix). \\
First, using Eq.(\ref{4.1}), 
we recalculate the asymmetry measured by SMC \cite{SMCp}
and E143 \cite{E143p} on the proton and deuteron targets at $Q^2= 10~
{\rm GeV}^2$ (SMC) and $3~ {\rm GeV}^2$ (E143), which are average $Q^2$ of
these experiments respectively. Obtained values of $\int g_1(x) dx$ through
the measured $x$ ranges are shown in the Table 1.\\
To get the values of the first moments $\Gamma_1^{p(d)}$ we
estimate unmeasured regions of SMC and E143 using their original
machinary.
Our estimations coincide with original ones except to the results in
small $x$ region unmeasured by SMC. We obtain the following results for
central values of
$\Delta \Gamma ^{p,d} = \int^{0.003}_0 g_1(x) dx$ at $Q^2=10 GeV^2$:
$\Delta \Gamma ^{p} = 0.003$ and $\Delta \Gamma ^{d} = 0.0022$, which
are smaller then the corresponding SMC estimations: 
$\Delta \Gamma ^{p} = 0.004$ and $\Delta \Gamma ^{d} = 0.0028$.
The errors coincide with ones cited in \cite{SMCp}.
The E143 estimations for $ \int^{0.029}_0 g_1(x) dx$ are not changed
because $Q^2$-evolution of the asymmetry is negligible at $x \sim 0.03$.
Results on the $\Gamma_1$ values are shown also in the Table 1.\\
We
would like to note that the E143 and SMC machinary may lead to 
underestimation of
$g_1^{p,d}(x,Q^2)$ at small $x$ and, hence, to underestimation of 
$\Delta \Gamma ^{p,d}(Q^2)$ (see the careful analysis in first paper in 
ref. \cite{GRSV}). Unfortunately, our procedure is not at work
 at $x \leq 0.01$
and we cannot check the SMC and E143 estimations of unmeasured 
regions here. To clear up
this situation it is necessary to add a careful small $x$ analysis to this
consideration that is a subject of our future large article \cite{KNP}.\\

\hskip -.56cm
{\bf Table 1.}~~ The first moment values of $g_1$ of the proton and deuteron.\\
\vskip -.9cm
\begin{table}[h]
\begin{center}
\begin{tabular}{|c|c|c|c|c|c|}
\hline \hline
                     &         &             &          &                 &  \\
$x_{min} -- x_{max}$ & $<Q^2>$ & target      & $\int_{x_{min}}^{x_{max}} g_1 dx$ &
$\Gamma_1$ & experiment \\
                     &         & type        &          &                 &  \\
\hline \hline
.003 -- 0.7 & $10~{\rm GeV}^2$ & proton      & 0.130 & 0.134 $\pm$
0.011 &SMC \\
.003 -- 0.7 & $10~{\rm GeV}^2$ & deuteron    & 0.038 & 0.036 $\pm$
0.009 &SMC \\
\hline
.029 -- 0.8 & $3~{\rm GeV}^2$ & proton      & 0.123 & 0.130 $\pm$
0.004 &E143 \\
.029 -- 0.8 & $3~{\rm GeV}^2$ & deuteron      & 0.043 & 0.044 $\pm$
0.003 &E143\\
\hline
\hline
\end{tabular}
\end{center}
\end{table}
\hskip -.56cm
As the last step we calculate the difference which is predicted by the
Bjorken sum rule $\Gamma_1^p - \Gamma_1^n$ :
$$\Gamma_1^p - \Gamma_1^n = 2 \Gamma_1^p -
2 \Gamma_1^d / (1-1.5 \cdot \omega_D),$$ where
$\omega_D=0.05$ \cite{SMCp,E143p} is the probability of the deutron to
be in a D-state .\\

At $Q^2=10~ {\rm GeV}^2$ we get the following results:
\bea
\Gamma_1^p - \Gamma_1^n &=& 0.190 \pm 0.038~~~~~~~~~~~~~~
\eea
to be compared with the SMC published value
\bea
\Gamma_1^p - \Gamma_1^n &=& 0.199 \pm 0.038~~~~~~~~~~~~~~ \mbox{(SMC
  \cite{SMCp})}
\nonumber
\eea
and the theoretical prediction
\bea
\Gamma_1^p - \Gamma_1^n &=& 0.187 \pm 0.003~~~~~~~~~~~~~~ \mbox{(Theory)}
\nonumber
\eea

At  $Q^2=3~ {\rm GeV}^2$ we get for E143 data:
\bea
\Gamma_1^p - \Gamma_1^n &=& 0.165 \pm 0.026~~~~~~~~~~~~~~
\eea
to be compared with
\bea
\Gamma_1^p - \Gamma_1^n &=& 0.163 \pm 0.026~~~~~~~~~~~~~~ \mbox{(E143
  \cite{E143p})}
\nonumber \\
\Gamma_1^p - \Gamma_1^n &=& 0.171 \pm 0.008~~~~~~~~~~~~~~ \mbox{(Theory)}
\nonumber
\eea
Note that only the statistical errors are quoted here.
To the considered accuracy they coincide with the errors
cited in (\cite{SMCp,E143p}). The above cited theoretical
predictions for the Bjorken sum rule have been computed in
\cite{LV} to the third order in the QCD $\alpha_s$.\\

As a conclusion, we would like to note
\begin{itemize}
\item The value of $\Gamma_1^p - \Gamma_1^n$ obtained in our analysis
is in the best agreement with the Bjorken sum rule prediction.
\item The values of $\Gamma_1^p$ and $ \Gamma_1^n$ themselves
obtained here do not change essentially.
The improvement for the Bjorken
sum rule is the result of the opposite changes of the $\Gamma_1^p$
and $ \Gamma_1^n$ values, when Eq.(\ref{4.1}) is used.
\item our observation that function $A_1^*(x)$ is $Q^2$ independent at
large and intermediate $x$
is supported by good agreement
of present analysis with other estimations \cite{ANR,GS95,GRSV} of
the $Q^2$ dependence of the $A_1$. A detail analysis will be present
later in the separate large article \cite{KNP}.
\end{itemize}
\hskip -.56cm

{\large \bf Acknowledgements}\\

\hskip -.56cm
We are grateful to W.G.~Seligman for
providing us the available CCFR data of Ref.\cite{CCFR},
A.V.~Efremov for discussions and to anonimous Referee for
critical notices lead to essential improvement of the article.\\
We also grateful to G.~Ridolfi for the possibility to present the
short version \cite{KP} of this letter on polarized structure function
 section of the Workshop DIS96.\\

\hskip -.56cm
This work is supported partially by the Russian Fund for Fundamental Research,
Grant N 95-02-04314-a.\\

\hskip -.56cm
{\large \bf Appendix}\\

\hskip -.56cm
The parametrization is used for CCFR data \cite{CCFR} :
\bea
xF_3(x,Q^2) = F_3^a  \cdot { \biggl( {
log(Q^2/\Lambda ^2) \over log(Q^2_0/\Lambda ^2) } \biggl) }^
{F_3^b},
\nonumber
\eea
where
\bea
F_3^a = x^{C_1} \cdot (1-x)^{C_2} \cdot
\biggl( C_3+C_4 \cdot (1-x) +C_5 \cdot
(1-x)^2+
\nonumber \\
C_6 \cdot (1-x)^3 +C_7 \cdot (1-x)^4 \biggr) \cdot
\biggl[ C_8+C_9 \cdot x+C_{10} \cdot x^2 + C_{11} \cdot x^3 \biggr]
\nonumber
\eea
\bea
F_3^b = C_{12}+C_{13} \cdot x+{C_{14} \over x+C_{15}}
\nonumber
\eea
and $Q^2_0 = 10~ {\rm GeV}^2$, $\Lambda = 200~{\rm MeV}$.\\
\vskip .2cm \hskip -.56cm
{\bf Table 2.}~~ The values of the coefficients of CCFR data parametrization.\\
\vskip -.9cm
\begin{table}[h]
\begin{center}
\begin{tabular}{|c|c|c|c|c|}
\hline \hline
 $ C_1$ &  $ C_2$ &  $ C_3$ &   $ C_4$ &  $ C_5$ \\
\hline
 0.8064 & 1.6113  & 0.70921 &  -2.2852 & 1.8927  \\
\hline \hline
 $ C_6$ &  $ C_7$ &  $ C_8$ &   $ C_9$ &   $C_{10}$ \\
\hline
 6.0810 &  4.5578 &  0.7464 &  -0.3006 &  3.9181\\
\hline \hline
 $C_{11}$ &   $C_{12}$ &   $C_{13}$ &  $C_{14}$ &   $C_{15}$ \\
\hline
-0.1166 &  10.516 & -5.7336 & -37.114 & 3.7452 \\
\hline \hline
\end{tabular} \end{center} \end{table}

%

%

\end{document}